\documentclass[sigconf,authorversion]{acmart}
\AtBeginDocument{%
  \providecommand\BibTeX{{%
    \normalfont B\kern-0.5em{\scshape i\kern-0.25em b}\kern-0.8em\TeX}}}

\copyrightyear{2023}
\acmYear{2023}
\setcopyright{acmlicensed}\acmConference[CHIIR '23]{ACM SIGIR Conference on Human Information Interaction and Retrieval}{March 19--23, 2023}{Austin, TX, USA}
\acmBooktitle{ACM SIGIR Conference on Human Information Interaction and Retrieval (CHIIR '23), March 19--23, 2023, Austin, TX, USA}
\acmPrice{15.00}
\acmDOI{10.1145/3576840.3578316}
\acmISBN{979-8-4007-0035-4/23/03}

\newcommand{\ah}[1]{#1}
\newcommand{\me}[1]{#1}

\begin{document}

\title{Much Ado About Gender}
\subtitle{Current Practices and Future Recommendations for Appropriate Gender-Aware Information Access}

\author{Christine Pinney}
\email{christinepinney@u.boisestate.edu}
\orcid{0000-0002-7791-4754}
\author{Amifa Raj}
\email{amifaraj@u.boisestate.edu}
\orcid{0000-0002-8874-6645}
\affiliation{
  \institution{People \& Information Research Team\\
  Boise State University}
  \streetaddress{1910 University Drive}
  \city{Boise}
  \state{Idaho}
  \country{USA}
  \postcode{83725-2055}
}

\author{Alex Hanna}
\email{alex@dair-institute.org}
\orcid{0000-0002-8957-0813}
\affiliation{%
  \institution{DAIR Institute}
  \country{USA}}

\author{Michael D. Ekstrand}
\email{ekstrand@acm.org}
\orcid{0000-0003-2467-0108}
\affiliation{%
  \institution{People \& Information Research Team\\
  Boise State University}
  \streetaddress{1910 University Drive}
  \city{Boise}
  \state{Idaho}
  \country{USA}
  \postcode{83725-2055}
}
\renewcommand{\shortauthors}{Pinney et al.}

\begin{abstract}
  Information access research (and development) sometimes makes use of
gender, whether to report on the demographics of participants in a user
study, as inputs to personalized results or recommendations, or to make
systems gender-fair, amongst other purposes. This work makes a variety
of assumptions about gender, however, that are not necessarily aligned
with current understandings of what gender is, how it should be encoded,
and how a gender variable should be ethically used. In this work, we
present a systematic review of papers on information retrieval and
recommender systems that mention gender in order to document how gender
is currently being used in this field. We find that most papers
mentioning gender do not use an explicit gender variable, but most of
those that do either focus on contextualizing results of model
performance, personalizing a system based on assumptions of user gender,
or auditing a model's behavior for fairness or other privacy-related
issues. Moreover, most of the papers we review rely on a binary notion
of gender, even if they acknowledge that gender cannot be split into two
categories. We connect these findings with scholarship on gender theory
and recent work on gender in human-computer interaction and natural
language processing. We conclude by making recommendations for ethical
and well-grounded use of gender in building and researching information
access systems.
\end{abstract}

\ccsdesc[500]{Social and professional topics~Gender}
\ccsdesc[500]{Information systems~Information retrieval}

\keywords{information access, gender, auditing, systematic review}



\maketitle

\section{Introduction}
Research and development of \textit{information access systems} \me{(IAS)} --- search engines, recommender systems, and similar systems that facilitate access to information\me{, often studied in conferences on information retrieval (IR) and related topics such as recommendation and user modeling} --- often engage with gender in some way or another.
These uses vary, from reporting the demographic distribution of participants in a user study to using gender as a feature in personalized results to seeking to ensure the system treats users or content providers of various genders fairly, among other objectives.
There has been little explicit consideration in this literature, however, about how gender should be used in information access.
Most work takes gender as a categorical feature that can be obtained from users \ah{or inferred from} the underlying data set and uses it as any other feature in the system.
There are several important questions about the use of gender in information access research, including: 

\begin{itemize}
\item When should gender be used, and when is it inappropriate, unhelpful, or harmful to use gender in research or practice?
\item When it is appropriate to use gender, how should gender be defined and operationalized?
\item Where and how should gender data be obtained? Are there methods that are best avoided?
\end{itemize}

Our goal in this paper is to document the current state of research practice with respect to these questions and provide a foundation for discussion, further research, and well-grounded practice among information access researchers, practitioners, affected parties, and others that moves the community towards thoughtful, principled use and non-use of gender.
We agree that it is indeed crucial for
search engines, recommender systems (RS), and other information access systems to provide effective, appropriate, and useful results to users of all genders and other demographic affiliations.
We argue that this is best done through careful attention to the meaning of gender and how its use and operationalization affects the people the system is aiming to assist, particularly people with marginalized gender identities and adverse experiences with computational and datafied representations of gender.

To that end, we organize this paper in two parts. First, we provide a systematic review \ah{and analysis} of \ah{the} use of gender \me{in recent publications in} key
information access research venues. \ah{We then} identify goals for which gender is used, ways it \ah{is} encoded, and \ah{the} data sources used to obtain gender information for users, content providers, and other affected people.
\me{Finally,} we build on this survey and relevant literature from other domains to provide recommendations for improving research and implementation practices around gender in information access.

We are certainly not the first to question how gender is used in computing systems. \citet{Hamidi2018-pt} and \citet{Scheuerman2020-dq} have done crucial work on the (mis)use of gender in human-computer interaction, and \citep{Devinney2022-pb} have looked at how it is used in natural language processing (NLP) research. \ah{This highlights how this issue is not unique to \me{IAS}; indeed, this is a common issue in quantitative social sciences writ large \citep{Westbrook2015}}. We complement their work \me{by} specifically \me{investigating} information access applications, including search and recommendation.

\section{Motivating Vignettes}
The use of gender as a variable in information access systems may be becoming more ubiquitous. Gender may be used as an input to a recommender system or information retrieval model. Some of the uses of gender may present themselves as more insidious than others. To motivate our interest in understanding the use of gender, we present two vignettes.

In China, Kentucky Fried Chicken partnered with Baidu to offer a product which provided food recommendations based on details inferred from a customer’s face at 300 stores in Beijing \citep{Etherington2016-ru}. In addition to inferring gender, the facial analysis product also inferred age and ``beauty'' \citep{Hawkins2017-vn}. The tool recommends different meals which are seemingly based on these factors. For instance, the author of the Guardian article was read as a woman in her 30s, and the system recommended a chicken hamburger meal. A press release from Baidu suggested that ```a male customer in his early 20s' would be offered `a set meal of crispy chicken hamburger, roasted chicken wings and coke', while `a female customer in her 50s' would get a recommendation of `porridge and soybean milk for breakfast'.'' 

\ah{Gender itself is inferred \me{in this system} from gender expression, which has been criticized in the literature which we discuss below. Moreover, strong assumptions are made about the role gender should play in product recommendation.} It’s not clear how, \ah{\textit{prima facie}, how} these meals correlate with these inferred features. \ah{In what way does it make sense for features such as inferred gender, beauty, or age \me{to} serve as a suggestion for meal items? Are those features indicative of purchasing behavior or desired products? To us, these features, inferred from personal appearance, make spurious product recommendations.} \ah{However, what we do know is that the system} presents a new avenue for massive collection of facial images and purchasing patterns, which could be used by Baidu to monetize other aspects of social and economic life in China.

Another, more positive, use of gender can be found in an audit conducted by Spotify to assess how female artists are represented and made visible to listeners through the platform’s discovery tools \citep{Epps-Darling2020-eh}. \ah{The authors of this study} found that recommendations had a slightly higher proportion of female artists than users' ``organic'' behavior \ah{(i.e. behavior which \me{was not} recommendation-driven)}, and further, that recommending more female artists correlated with increases in later user-initiated streaming of music by female artists. 

\ah{In this case, gender as a variable is used as an identifying feature of a recommended product.} Such work \ah{can be} valuable in understanding how information access technologies interact with societal discrimination, when they propagate such biases, and how they may be deployed as interventions to promote more equitable information economies \citep{fnt-fairness}. 

\section{Background}

\me{Gender has been discussed in various ways in information access research throughout the history of the relevant fields, and there is also a rich literature on the construct of gender and its interaction with data and computation.
To set the stage for our formal review in Section~\ref{sec:review}, we first briefly outline some of that background here.}

\subsection{The Uses of Demographics in \me{IAS}}
As noted in the introduction and explored much more thoroughly in our systematic review, there are a variety of ways that gender appears in information access research. One of the earliest recommender systems, Grundy \citep{Rich1979-eh}, explicitly used a user’s gender as a component of its model of their preferences and incorporated gender stereotypes into its initial recommendations (which the user could refine through subsequent conversational interaction); in modern personalization, gender is one of the many attributes data brokers routinely collect and sell to companies to use for a variety of purposes \citep{Crain2018-lo}. Early work on matrix factorization for collaborative filtering used a gender affinity axis (“geared toward females” vs. “males”) to illustrate the idea of embedding movies \citep{Koren2009-fo}. A more recent line of work seeks to understand information access systems’ differential impacts to see if they are treating people of different genders “fairly” as users \citep{Ekstrand2018-um, Mehrotra2017-ns}, as producers of the information being retrieved \citep{Ekstrand2021-iu, Epps-Darling2020-eh, Ferraro2021-cx}, or as the subjects of that information \citep{Ekstrand2022-ma, Karako2018-oz, Metaxa2021-jj}.

Aside from discussions in limitations sections of some of these papers, there is little work on when, why, and how gender is and should be used in information access research, or putting this work in the context of discussions about gender in social science or other computing fields such as human-computer interaction. This is the gap we seek to fill in this paper.

\subsection{Gender as a Category}
Much of the literature within sociology and gender studies has focused on the differences between gender and sex. Typically, ``sex'' is used to refer to biological characteristics while “gender” is related to internal perceptions of self and how external society sees individuals. However, gender and sex are entangled, and sex itself is socially constructed by scientists, policymakers, and technologists \citep{Fausto-Sterling2000-wa, Sanz2017-qi}.

Gender scholars, as well as transgender and queer activists, have also made the distinction between gender identity and gender expression. Gender identity typically refers to one’s own internal understanding of gender and self-identification. Gender expression refers to how one presents one’s own gender and wants to be seen by the world. These both can fit into binary notions of gender, but can also be expansive and encompass a constellation of different identifications and notions of what self-expression can entail. Moreover, gender expression can be broken up both internally (how one is expressing one’s gender and feels about it to themselves) and how others perceive that individual's gender (perceived gender expression). In this article, we follow \citep{Scheuerman2021-wm} and focus on discussing gender, given that technological artifacts and systems typically discuss social constructions of gender as datafied by informational systems. However, it is important to note that many types of information access systems may make claims about having data on sex (e.g. through medical imaging or genomics). 

Gender data can be obtained in a plethora of ways, depending on the modality. There is robust literature within social science research on how to survey for gender, especially when that measurement moves beyond the male/female gender binary. In survey research, much of the focus has centered around ensuring that population-level estimates can be inferred from a sample that is attentive to the individuals who do not fit into either the category of male or female. The Williams Institute has developed tools which ask respondents if they identify within the binary and then asks about transgender status \citep{The_GenIUSS_Group2014-ta}. This has been criticized as being too reductive, however, and may not be applicable for smaller scale studies. Others have focused on attempting to obtain a measure of how others may perceive their gender expression \citep{Magliozzi2016-yt}. \ah{More recently, many others have addressed how to approach gender as a matter of data justice using intersectional feminist and queer theory lenses \citep{d2020data, guyan2022}}.

Gender data come from many different places in the papers we examine, so we do not distinguish between gender identity and gender expression. However, it is important to note that these two categories are used nearly interchangeably in the computer science literature that we surveyed. 

\subsection{Gender in Computational Research}
With the rise in attention to facial recognition as a technology, many researchers within HCI and AI have focused on the attribution of gender to individual data traces, typically images of people. \citet{Keyes2018-gq} has written on the dangers of automatic gender recognition (AGR), \citet{Scheuerman2019-jr} have written on how AGR systems perform worse on trans and gender non-conforming people, and how these systems cannot legibly recognize non-binary people. Gender non-conforming and transgender individuals also feel as though these systems produce harm by involuntarily gendering them \citep{Hamidi2018-pt}. ``Gender'' is also necessarily raced; that is, binary genders themselves are the endpoint of processes of centuries of European colonization \citep{Scheuerman2021-wm} and erase other genders which were part of indigenous and non-Western societies. Moreover, gender assessments are typically not accorded the same status to non-white women, especially Black women, as evidenced by \citep{gender-shades}.

Moreover, although there is less academic research in the interaction of gender and text, this is still a strain of research which manifests in a few different registers. There is a body of work which attempts to predict gender from textual prose (e.g. \citep{Peersman2011-tq}, however much of the work in natural language processing focuses on the notion of gender bias in text and text representations. One of the most major of these interventions \citep{Bolukbasi2016-tl, Caliskan2017-ka} suggests that pre-trained embedding spaces exhibit sexist biases (e.g. doctors are to males, whereas nurses are to females). Recent work has suggested that, although there is significant work in gender bias in NLP, few of these papers engage with gender theory, consider non-binary genders, or consider the intersectional, already-racialized notion of gender \citep{Devinney2022-pb}.
At the intersection of computer vision and natural language processing, gender and racialized-gender bias persist in multi-modal domains, such as image search \citep{Noble2018-nc}, text-image benchmarks \citep{Diaz2021-fa}, and multi-modal models such as SCAN and CLIP \citep{Wang2021-qe}.

\section{Review of Current Practices}
\label{sec:review}

In order to better understand the landscape of the use of gender in information access systems, we conducted a survey of all papers which mentioned sex or gender in key information retrieval and recommendation systems publication venues. We desired to assess what, in particular, this academic community was doing with the concept of gender in academic outputs.

\subsection{Methods}
To collect a set of papers to analyze, we searched for all papers that mentioned gender-related words in SIGIR, CHIIR, RecSys, UMAP, and TOIS papers in 2017-2021 using the ACM Digital Library search.
\me{We selected these venues to furnish examples representative of multiple perspectives in, particularly from a computer science perspective; papers in these venues are influential across both research and practice.}
\ah{We selected these years because we wanted to take a snapshot of \me{relatively current} research within this field rather than attempt to make any larger claims about changes over time\me{, particularly extending to earlier days of information access research}.} 

We constructed a codebook based on a sample of articles matching our criteria. \ah{The codebook was constructed at the guidance of the third author, a sociologist who focuses on the intersections of technology, race, and gender, \me{and} the final author, a senior computer scientist focusing on information access systems. New questions were added as needed. For instance, we began the study focusing online on whether there was a gender variable and the goals of using gender, with the assumption that most articles would address user gender. However, we then came to understand that gender may have different referents (e.g. the data instance), and that there may be multiple referents. Moreover, we began to find that many of the uses of gender were part of an audit process to detect bias, so we added another question regarding those explicitly.} 

The lead author then coded for each of the variables in our codebook across all the articles. \ah{All authors met weekly to discuss the coding process and resolve ambiguities, and to work through exemplar cases with the lead author. Because a single person coded all the articles, we do not report interrater reliability metrics. } The full coding process, the codebook, \ah{and the dataset} are available in citation \citep{gender-data}.

\subsubsection{Variables}
For each paper, we coded for several different variables. \ah{Table~\ref{table:variables} provides a summary at a glance.}

\paragraph{What is the primary referent?}
The referent is the group of people who gender is being attributed to. This may be users of a recommender system, subjects of particular data instances (such as clothing or musical artists), or annotators who are labeling data. In cases in which the paper conducted a user study using a crowdworking platform, we coded study participants as ``users.''
\ah{While we began this study anticipating only ``users'', ``subjects'', and ``providers'' being our referent, we added annotators as we continued to code.}

\paragraph{Is there a gender variable?}
We determined if there was any kind of gender variable in the article text at all. If there is no gender variable, then this disqualifies answering other questions about the paper. We coded a paper as “applied” when the model or experiment in the paper did not use gender, but the authors suggest that gender could be used with their method.   

\paragraph{What are the gender categories?}
This variable outlined which values the gender variable will take. Sometimes these were explicitly mentioned, but often they will be obscured in a table or implicit in a statistical model. Moreover, we also noted if the authors coded for a third gender, such as ``other'' or ``non-binary.'' We also coded for whether the authors verbally acknowledged that gender was non-binary, but did not operationalize this in any way. We did so because we hypothesized that some authors would make a textual note that gender was non-binary, but then continue using binary values for gender.

\paragraph{How is gender determined?}
We coded how the authors are obtaining the gender label. The gender label itself may come from self-identification by the user, or from an inference being made by the authors, third-party annotators (such as crowdworkers), or an automated system. We began from two expected categories (self-identification and machine-inferred) but added \ah{crowdworker inference as we noticed this in the data.}
\paragraph{Is this paper about bias and/or fairness?}
Many papers will be about assessing the bias with a particular system or dataset, attempting to debias a dataset, or create a fair dataset or method. This would be more akin to the auditing example noted above.

\paragraph{Goals of using gender}
Lastly, we coded for the ``goal'' of the use of the gender variable. Instead of defining a set of discrete goals which gender was used for, this was an inductive category, in which we added different goals progressively. This included some goals which we expected at the start of the research project, such as ``Personalize based on gender`` or ``Gender prediction`` (both used in the KFC China example above), but also encompassed some surprising uses. We discuss these inductively coded goals below in the findings.

\begin{table}
    \begin{tabular}{|r|l|l|}
        \hline
        \bf{Variable} & \bf{Possible values} \\
        \hline
        Primary referent & User/Subject/Provider/Annotator \\
        Gender variable? & Yes/No \\
        Gender categories & Binary/Binary+Other \\
        Gender determination & Self-identification/Annotator/Inferred \\ 
        Bias/fairness? & Yes/No \\
        Goals                 &  User study/survey \\ 
                              &  Gender personalization \\ 
                              &  Audit system behavior \\ 
                              &  Gender prediction \\ 
                              &  Protect gender variable \\
                              &  Persona generation \\
        \hline
    \end{tabular}
    \caption{Summary of variables}
    \label{table:variables}
\end{table}

\subsection{Overview of Data and Univariate Findings}
We collected 801 papers from 4 conferences (CHIIR, SIGIR, RecSys, and UMAP) and one journal (TOIS); of these, we coded 598 papers and excluded 203 workshop summary papers that didn’t have sufficient peer review to code. Of the 598 coded papers, we found that 73 papers had a gender variable of interest, 442 did not have a gender variable, and 57 had a gender variable that was “applied.”

\subsubsection{Gender Referent}
In each paper, the authors attribute gender to a specific object — the person or thing that the authors are referring to when discussing gender. If authors attributed gender to multiple entities, one entity was labeled as the primary referent and the paper was coded as having multiple referents. We identified 4 types of referents with which authors associated gender.

\paragraph{User Referent (52 papers)}
This set of papers considers gender association of users who interact with systems \citep{Cheng2017-op, Hashemi2017-sy, Marlow2017-tz, Ramos2020-jk, Tavakol2020-nl}. This user interaction may be direct where gender identity is self-declared (user study or survey), or it may be indirect where gender identity is annotated or inferred (annotation of user-generated profile, facial inference). For example, \citet{Rozen2021-eh} used user-stated gender information to evaluate their proposed system in predicting user demographic attributes, namely gender, from user browsing data and generated comments on news articles.

\paragraph{Subject Referent (15 papers)} 
In this group of papers, gender is associated with subjects or items. Gender of items can be inferred from item content, for example, song lyrics, documents, and dataset labels \citep{Nasirigerdeh2021-rp, Barman2019-ez,
Solomon2018-sw, Yeshambel2021-nq}. For instance, \citet{Rekabsaz2020-sz} use gendered keywords to identify female/male magnitude of retrieved documents and provide metrics for measuring gender bias in retrieval sets. They use an annotated dataset of gendered and non-gendered queries to demonstrate the use of these metrics in measuring gender bias of a result set. 

\paragraph{Provider Referent (5 papers)}
Items can be associated with the gender of item providers or content creators (music artists, book authors), so the gender of the providers or creators is often assigned to the items \citep{Agosti2019-pa, Ekstrand2018-um, Mukherjee2020-oh}. For example, \citet{Ferraro2021-cx} identify gender bias of artists in music recommendations and propose a progressive re-ranking method that achieves improved gender balance of musical creators in recommendation systems.

\paragraph{Annotator Referent (1 paper)}
This type of paper refers to the gender of the annotators where their act of annotation is significant (compared to if they serve as test users). In the single paper in this category \citep{Zhan2019-br}, the authors collected annotators' gender information to develop noise-aware sentiment classification models and illustrate the possible effect that demographic attributes may have on an annotator’s response. 

\subsubsection{Gender Determination}
During the coding process, we identified several ways with which authors determined the gender of the referent(s). The majority of papers (68) involved one method of gender determination, but five papers used two.

\paragraph{Self-identification (53 papers)} 
In these papers, gender is determined with self-declarations of gender identity. In some cases, users declare their own gender (among other demographic attributes) while participating in a study or while using a system \citep{Braslavski2018-qv, Kleinerman2018-lk, Zhan2019-br}; in others, authors use publicly available datasets that provide demographic data where it can be assumed that gender was self-declared \citep{Tian2020-rq, Sanchez2019-xg, Qu2018-in}.

\paragraph{Annotators (16 papers)}
In this work, human annotators assign gender for users, providers or subjects \citep{Barlas2019-dg, Han2018-dc, Song2018-yt}. In \citep{Ekstrand2018-um}, the authors use a dataset where the gender of book authors was annotated by library professionals.

\paragraph{Inferred (7 papers)}
In these papers, gender is interpreted from item content, users’ personal information, interaction behavior or with the help of annotators. We identified papers that use users or providers name, voice, and images for inferring gender \citep{Ahmadvand2020-wu, Ghosh2021-jj, Marlow2017-tz}. For example, \citet{Mukherjee2020-oh} use a gender identification tool that infer users’ gender from their username and country of origin.

\subsubsection{Categories of Gender}

\paragraph{Binary (62 papers)}
This group of papers considered gender as a binary variable where they categorized gender into men and women. This is regardless of referent or determination type. These papers also do not acknowledge that gender is non-binary \citep{Ahmadvand2020-wu, Braslavski2018-qv, Mukherjee2020-oh, Otterbacher2018-kl, Salminen2018-ml, Wu2020-gm}. 

\paragraph{Acknowledgement of non-binary gender, or the use of a third gender category (11 papers)}
The other eleven papers consider the concept of gender beyond binary categorization. In six of them, the gender categories were extended to include unisex, mix-gender, and non-gendered groups. For instance, in \citep{Eshel2021-cx}, the authors considered unisex and mix-gender categories along with men’s and women’s categories to predict buyer’s size preference in e-commerce.  The remaining four papers acknowledged the limitations of representing gender as a binary construct but continued to do so in any case. For example, in \citep{Ekstrand2018-um}, the authors use a binary gender variable to assess the results of collaborative filtering methods in book ratings and recommendations with respect to the gender of content creators, but include discussion of the negative effects and consequences of representing gender as binary. 

Notably, none of these papers provided classifications which affirmed non-binary gender identities. This is distinct from papers which provide a ``non-gendered'' categorization, such as ``unisex'' or ``other'', as noted from the examples given in the prior paragraph. We discuss positive examples of affirming non-binary gender identities in the discussion.

\subsubsection{Bias and Fairness}
With the rise in the interest of bias, fairness, and ethics in machine learning systems, and the development of new venues such as FAccT/FATML, a concomitant rise has been seen in the interest in the information access space. We coded for whether the papers dealt with issues of bias or fairness in IR systems. Of the 73 coded papers, nearly one-third (24) were concerned with bias or fairness. 

\subsubsection{Purposes and Uses of Gender}
We used an inductive coding method to assess the goal of using a gender variable. Inductive coding is typically used in grounded theory methodology \citep{Charmaz2006-ym} in which one does not presume a set of categories on some type of text, such as an interview transcript\me{; we wanted to understand the types of goals directly from the literature instead of imposing our assumptions on it}. In this case, we focused on the paper overall, rather than doing line-by-line codings.

By “goal”, we refer to the intention or technical achievement attempted by the method with respect to the gender variable. This is often, but not always, distinct from the goal of the paper itself. As an example, a paper which attempts to develop a state-of-the-art collaborative filtering recommender system with demographic data as a goal may integrate a gender variable as part of a vector of demographic features. In this case, the goal would be \textit{Gender Personalization}.

There may also be the cases in which the gender variable is used towards some other, broader end. For instance, a paper which attempts to show how errors of demographic inference get propagated in a fair ranking system would be characterized as “auditing system behavior,” but not “gender prediction.”

We developed ten distinct purposes of gender. The majority of papers (57) were labeled with one code but a handful (16) were coded with two or three codes. The ten purposes are outlined below.

\paragraph{User Study or Survey (31 papers)}
In this group of papers, users are asked to participate in a user study or are respondents in a survey where they assess model outcomes and provide feedback on a subjective aspect of a system. User responses are analyzed for measurements of perceived usability (user perception, user behavior, user knowledge retention). In this case, gender is often collected as a salient feature among other demographic features (age, location). For instance, in \citep{Otterbacher2018-kl}, the authors provided participants with a set of questions pertaining to a gender-biased result set of images to measure their perceived bias and search engine objectivity. In their assessment, the authors collected demographic information including gender, and determined measurements of two types of sexism detected in users in order to analyze the effect of a user’s sexist biases on user perception of gender bias in image retrieval.  

\paragraph{Gender Personalization (21 papers)}
In this group of papers, the authors use gender as part of a user profile to personalize recommendations. For instance, in \citep{Cheng2017-op}, the authors utilized user-specific information (gender, age, social status) to improve musical artist recommendations and to assess long-term music interests of users. 

\paragraph{Audit System Behavior (20 papers)}
In this genre of papers, the authors evaluate the behavior and outcomes of an existing model or framework and offer recommendations regarding functionality and/or fairness based on analysis results. Gender is highlighted among other demographic features both in the datasets used and when assessing results for fairness. For instance, in \citep{Rekabsaz2020-sz} the authors generated a dataset of non-gendered queries as input for several neural ranking models and measured the resulting gender bias.  

\paragraph{Gender Prediction (7 papers)}
In these papers, the authors infer a gender variable from existing data instances and typically use them towards some other system end, such as improving the personalized recommendations. For instance, in \citep{Sheikh2019-pq}, the authors utilized a deep-learning collaborative filtering approach to better predict size and fit of users within an e-commerce platform. To address the issue of data sparsity on user-item interactions, their model learned latent representations and implicit features of users (age, gender).

\paragraph{Protect Gender Variable (3 papers)}
In this group of papers, the main focus is privacy protection around a set of demographic variables, of which gender is highlighted. The authors often first simulated the system or model’s behavior to illustrate privacy violations and/or data leakage. To counteract the issue, the authors then proposed and demonstrated an adversarial method designed to mitigate privacy leakage and provide better protection for users’ sensitive attributes, namely gender. For instance, in \citep{Leiva2021-lk}, the authors demonstrated the relative ease with which user behavioral data can be unobtrusively retrieved during web browsing via mouse cursor movements and subsequently used to predict demographic attributes (age, gender). They then provided a web browser extension that implements their proposed mitigation technique to obfuscate user demographics.

\paragraph{Persona Generation (3 papers)}
This genre of papers specifically analyzes user perceptions of profile representations derived from user data. The authors collected demographic data (age, gender) from participants and assessed the design of automatically-generated personas with respect to participant responses. In this way, gender is highlighted as a demographic point of interest in both users and user perceptions of gendered personas. In \citep{Salminen2019-sb}, for instance, the authors conducted a survey measuring user perceptions of pseudo-personas, specifically in response to pairs of identical profiles where the profile features a smiling picture versus a non-smiling picture. They found gender to be an influential attribute of generated personas, wherein variation in the gender of participants resulted in perceptual variation of the gendered personas. 

\paragraph{Indexing Clinical Trials (2 papers)}
In this genre of papers, the authors evaluate query expansion and reduction techniques and work to determine optimal feature configurations to improve information retrieval within the medical field. The authors utilize a gender variable (among other demographics) to improve query results. In \citep{Agosti2019-pa}, for instance, the authors evaluated a precision medicine search engine and its functionality in retrieving scientific literature and clinical trials in which they employ four steps: an indexing step, a query reformulation step, a retrieval step, and a filtering step. In the indexing step, the authors included a gender field (among other demographic fields) to index clinical trials and used these fields to determine eligibility in the filtering step.

\paragraph{Gender Diversity \& Inclusion (1 paper)}
In this body of papers, the methods involve using gender, amongst other demographic attributes, to algorithmically determine diversity and inclusion in model outputs or a UX surface. In the single example in our dataset \citep{Mukherjee2020-oh}, the authors offered an unsupervised summarization framework that provides a user with control over the shape and content (e.g., the gender of reviewers) of aspect-based summaries of tourist reviews on TripAdvisor. 

\paragraph{Linguistic Gender (1 paper)}
This set of papers deal with how to negotiate gendered aspects of language, including pronouns, nouns, and other gendered components. In our single example \citep{Yeshambel2021-nq}, the authors morphologically annotated Amharic (a gendered language) for the purpose of extending the application of lexical analysis to include more languages. 

\paragraph{Gender Interest Personalization (1 paper)}
In this last group of papers, they deal with dyadic gender preferences, rather than the gender of the referent themselves, which would fall under the concept of Gender Personalization. In our sole example \citep{Makhijani2019-op}, the authors focused on a dating app context where “match” suggestions depend upon the user’s specific gender preferences of prospective companions.

\subsection{Bivariate Analysis}
The prior section provided an overview of our data findings for each of the respective variables we coded for in our review of papers. In this section, we dig into some of the trends of data across time and variables.

\subsubsection{Time Trends}
\begin{figure}[tb]
\includegraphics[scale=0.20]{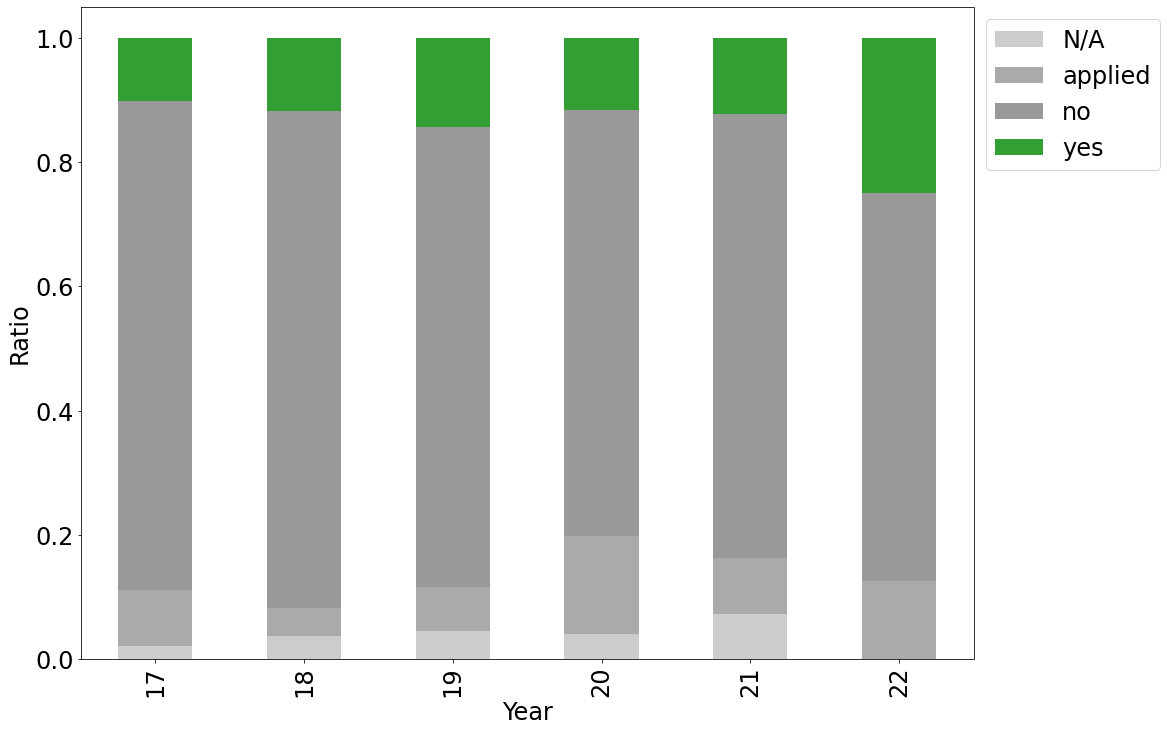}
\caption{Breakdown of whether the paper had a gender variable by year. “N/A” is used when papers refer to phrases such as “sexuality” and not biological sex.}
\label{fig:gender-year}
\end{figure}

Figure~\ref{fig:gender-year} shows the breakdown of our dataset by year. There has not been more of a focus on gender across time. There is a slight increase in the number of papers which mention a gender term, but about the same proportion of papers contain a gender variable from year to year. However, there are some notable changes across the goals of the use of a gender variable across time.

\begin{figure}[tb]
\includegraphics[scale=0.17]{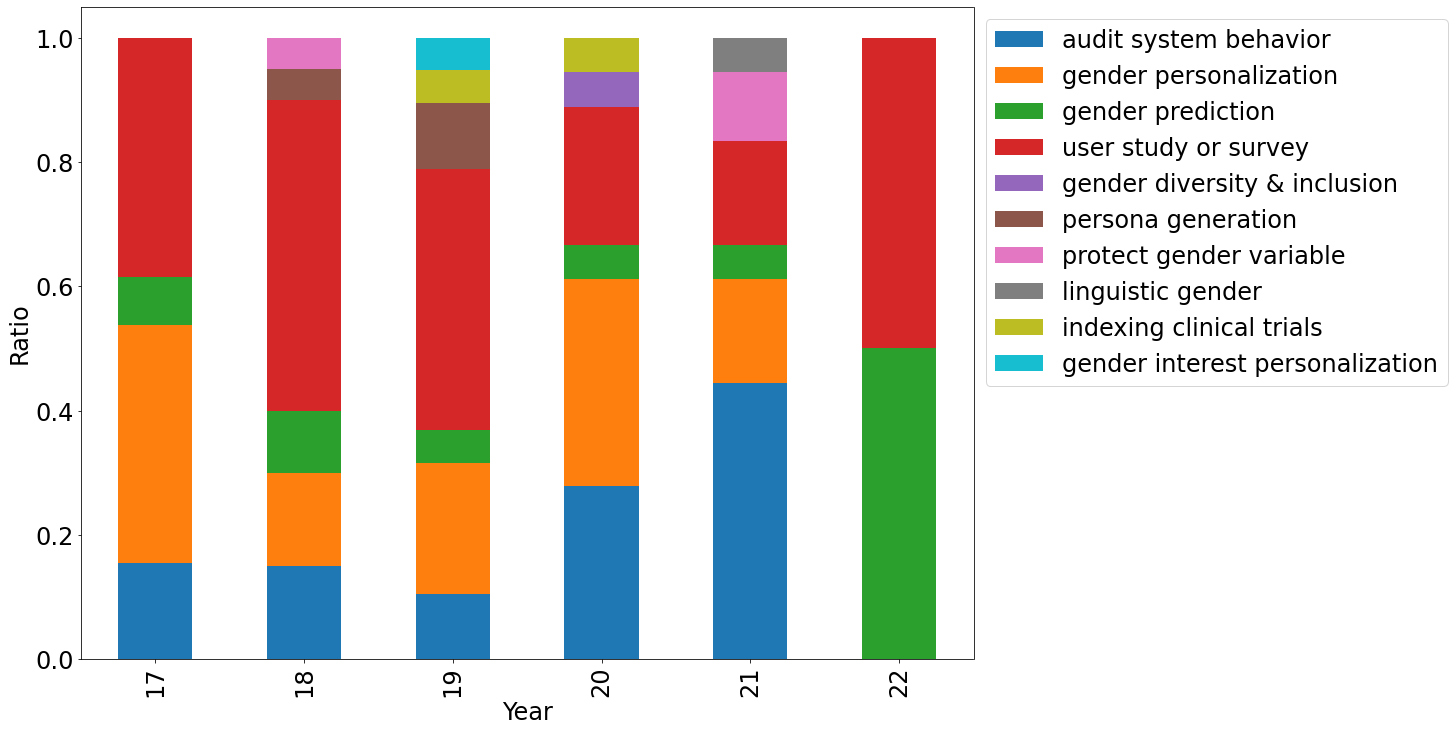}
\caption{Goals across time}
\label{figure:goals-year}
\end{figure}

The goals of using a gender variable have changed across time. The top two goals (``user study or survey'' and ``gender personalization'') are somewhat persistent across the study period, with the prior category peaking in 2018 and the latter in 2020. However, our third most prevalent category (“audit system behavior”) has been steadily climbing since the beginning of the study period, with its peak in 2021. 

\me{Similarly}, the use of a gender variable with the intent of assessing or testing for some kind of bias or fairness issue has risen across time, from two papers in 2017 to eight papers in 2021. In fact, in 2021, the majority of papers (8 of 15) dealt with fairness issues.

\subsubsection{Bias and Fairness}

\begin{figure}
\includegraphics[scale=0.175]{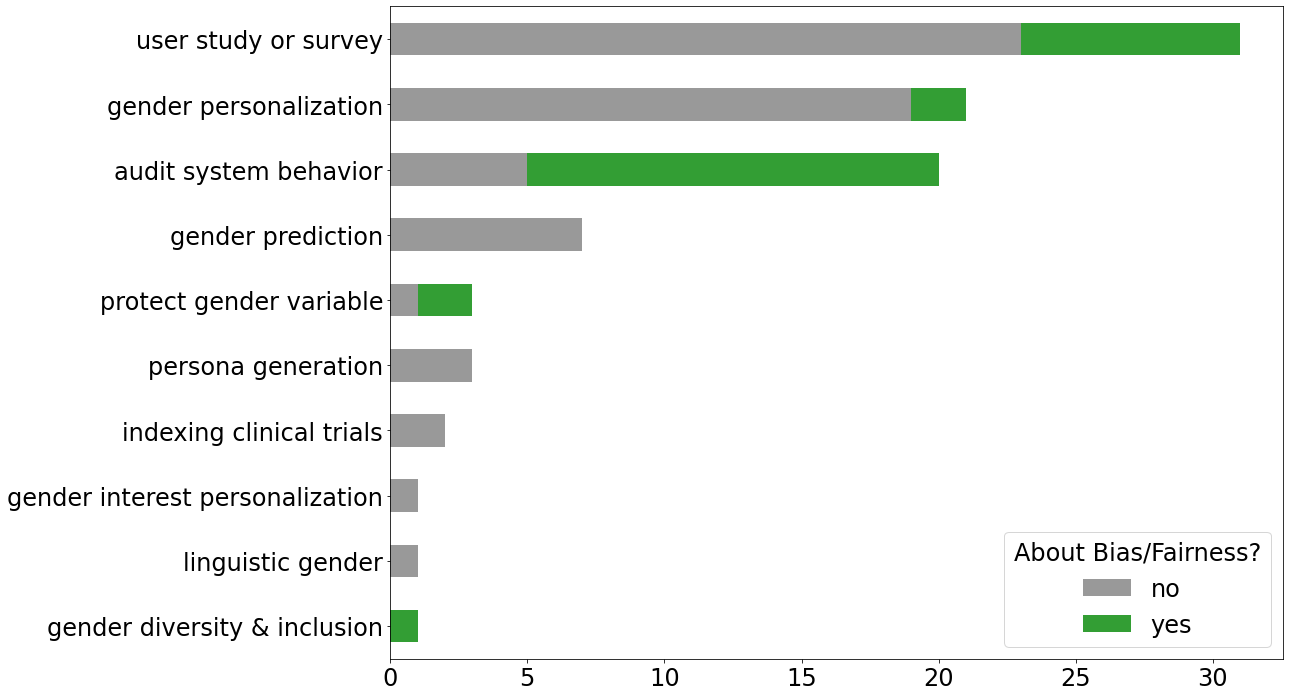}
\caption{Breakdown of bias and fairness by goal.}
\label{figure:bias-goals}
\end{figure}

When looking at the evaluation of bias and fairness as it pertains to each individual goal (Figure~\ref{figure:bias-goals}), we have found papers which audit system behavior address this topic significantly more often than papers which use gender variables towards any other end. User studies and surveys address bias at the second highest rate. None of our coded papers with the goal of ``gender prediction'' address bias or fairness in their discussion, and only two of those papers with the goal of ``gender personalization'' make note of this topic. The one paper which addresses gender diversity and inclusion deals with fairness issues.

\begin{figure}
\includegraphics[scale=0.175]{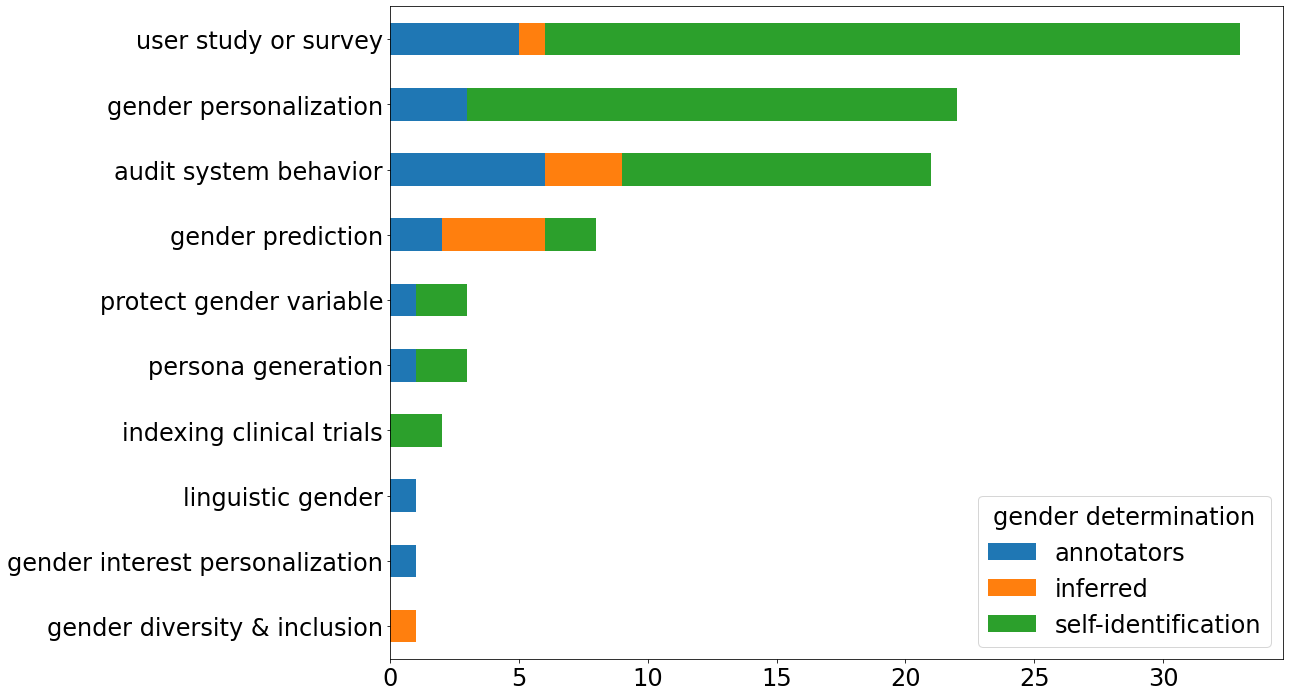}
\caption{Gender determination by different goals of the gender variable.}
\label{fig:determination-goals}
\end{figure}

\subsubsection{Gender Determination and Goals}
Figure~\ref{fig:determination-goals} shows the bivariate relationship between gender determination and paper goal. Most of the papers used ``self-identification'' as a gender determination. This is overwhelmingly the case for user studies (27), gender personalization (19), and auditing of system behavior (12). However, only two papers with the goal of ``gender prediction'' use ``self-identification'' as a gender determination, whereas all but two papers regarding ``gender personalization'' use “self-identification” over both inference and annotation. Significantly, papers which do gender prediction mostly use an inferred gender, which is not surprising, given the method. However, three papers which audit a system’s behavior use inferred gender, and one uses it in the case of user studies.

\subsection{Discussion}
From our analysis, there are several areas worth noting with regards to the use of a gender variable. Most notably, we found  no positive incorporation of non-binary genders within the papers we reviewed: that is, no papers successfully affirmed or accounted for non-binary gender identities. Although there a small portion of papers provided additional categories of gender beyond the binary male and female labels, it is important to note that the absence or neutrality of gender (as implied by “unisex” and “non-gendered” classifications) is not synonymous with non-binary gender identities. Over time, it appears that discussion, or, at the very least, acknowledgement of gender as non-binary has increased, but the successful utilization of a non-binary gender variable has yet to be made.

Secondly, there has been more awareness in fairness-oriented uses of gender variables in this research community, and it has gone up over time. Although there appears to be more of an effort on this front with goals like “audit system behavior”, it remains that papers with the goal of ``gender personalization'' and ``gender prediction'' fail to properly analyze the implications of their findings or model behavior in reference to gender bias and fairness. However, it may be the case that these two types of goals are antagonistic or fundamentally at odds with fairness and ethics, as suggested by \citet{Keyes2018-gq} and \citet{Scheuerman2019-jr}.

Third, the most frequent goal of using a gender variable is as input to an analysis in a user study or survey. This suggests that these authors are studying how differently gendered individuals respond to particular systems, which may be an encouraging result. More troubling, however, is the frequency at which systems attempt to personalize results based on gender. This itself makes major assumptions about what individuals may prefer, based on a gender variable, rather than on user preferences. We discuss alternative practices of personalization below. A more heartening development, however, is that auditing of system behavior has increased over the past five years, and that most of these studies do this with some kind of fairness evaluation in mind. 

Lastly, across all papers, gender self-identification is the norm, rather than the exception. Self-identification is the most ethical manner of collecting gender data, although the exact method of doing so is still an area of discussion and research, as noted in our literature review above. In a small number of cases, however, gender is inferred or labeled by third-party annotators. Third-party evaluations, either by crowdworkers, paper authors, or machines, may perpetuate gender stereotypes or be another vector of misgendering. When users self-declare their own gender identities within a dataset, they are less likely to be misgendered by a system or model using that data than when human annotators or systems infer gender identities from data traces, such as product selection, names, face images, or texts that the individual writes. Self-declaration of gender, however, does not foreclose the possibility of misgendering, because much self-identification data are collected with only binary gender categories built into the systems which collect these data in the first place.

\section{Recommendations}
Researchers and practitioners need to proceed with care in dealing with gender in computational research. Depending on the goal, use, and determination of gender, both the research process and findings of such research may be harmful. This harm may be direct, as when a system misgenders a person, or it may be indirect by handling gender in a reasonable way on its own but when combined with other downstream components causes harm. In this section, we provide some high-level recommendations and guidelines about using gender information in research on information access systems. We are not providing definite rules of using gender in computational research; rather, we are providing recommendations that researchers and practitioners can consider to avoid inappropriate use of gender in their work. We also expect future work to build on these guidelines as both understanding and technical possibilities evolve.

\subsection{When to Use Gender?}
Researchers should first determine whether it is appropriate to use gender in the first place. For some applications, contexts, or goals, using gender in some way may be beneficial; for others, it may just not be useful; and in a number of cases it is likely actually harmful.

Auditing system performance, particularly for fairness and equity concerns \ah{(the goal of 20 of 73 papers)} seems a relatively positive use of gender. Its purpose is to identify and mitigate gendered harms the system may inflict or reproduce, and the results are usually only made visible in aggregate (so errors in gender determination are rolled up in statistical aggregates, rather than present in a table of genders of individual people, although public datasets to support such audits do include individual-level gender annotations). For example, \citet{Ramos2020-jk} examined systems that rank people and may have reputational implications to ensure that the resulting reputation is independent of gender. Care is needed, though, in order not to undermine the fairness or equity goal: work that aims to improve fairness but only does so within a binary gender construct, for example, may reinforce discrimination against non-binary people. Moreover, audits of system behavior that infers gender on individuals may reproduce harm by guaranteeing that a system works only for individuals who conform to stereotypical gender presentations or expressions. \ah{Lastly, this work may be used to diversify information access systems (e.g. \citep{mitchell2020}), but the same caveats for doing so via gender inferrence remains.}

Overall, we advise against personalization based on gender as a goal or component of a system \ah{(the goal of 21 papers)}. Such personalization inherently depends on stereotypes about peoples' interests and capabilities, either existing stereotypes derived from societal assumptions or new ones derived from data. This contradicts the premise of online personalization based on extensive user profiles, as implemented in collaborative filtering, that we can personalize to a user’s particular needs and tastes rather than relying on unpersonalized or group-based assumptions. As \citeauthor{Riedl2002-bc} argued \citep{Riedl2002-bc}, recommender systems should ``box products, not people.'' The literature we have surveyed has not made a compelling case for gender-based personalization, but rather assumes that it is a reasonable thing to do or does it because it has been done before. There is also reason to be suspicious of using gender for personalization even in cold-start scenarios before individual user feedback is available: because the feedback from which personalized systems learn is not entirely exogenous, but is partly a response to the system’s previous outputs \citep{Chaney2018-us}, the system may learn future ``data-driven'' stereotypes not from organic user interactions but from its own initial assumptions. That is, if initial recommendations are derived from erroneous gender stereotype assumptions, data from the resulting interactions may reinforce those assumptions not because they are an accurate model of user interests, but because the user would have clicked on any comparable recommendation. Further study is needed to identify whether and to what extent this is happening, but it is a risk that should be taken seriously.

Lastly, following critical work on automated gender recognition \citep{Keyes2018-gq, Scheuerman2019-jr}, we also advise against gender prediction in information access systems \ah{(the goal of 7 papers)}. Many of the papers we find in our data focused on gender prediction aim to make that determination from user behavior, such as written internet text \citep{Rozen2021-eh} or more esoteric data such as spatial trajectories \citep{Solomon2018-sw}. However, similar to our warning against gender personalization above, these predictions may perpetuate gender stereotypes and re-entrench them by making those determinations based on data instances which bear no relationship to gender, and will most likely misrepresent individuals who are transgender or gender non-conforming. 

\subsection{How to Use Gender?}
If it is appropriate to consider using gender in some way, actually operationalizing and applying it requires additional careful consideration. In this section, we focus on more ethical goals of using gender and ethical strategies of gender determination.

Our first recommendation is to use an inclusive concept of gender to the extent possible. Restricting work to a male/female gender binary limits its applicability and reproduces exclusion of gender minorities. Data selection is the first obvious application of this principle, but it goes beyond simply the data; for example, while 
\citet{Ekstrand2021-iu} (expanding on \citep{Ekstrand2018-um}) acknowledged non-binary gender identities as valid and an important limitation, the metric and resulting statistical method they employed cannot be applied to non-binary attributes. Even when only binary data is available, we advise against methods that cannot be applied outside of binary contexts, so that the analysis can be updated if and when more inclusive data is located or produced \citep{Raj2022-js}.

Examples of inclusive gender data and analyses are rare, but the TREC Fair Ranking track and dataset \citep{Ekstrand2022-ma} does use non-binary gender identities for bibliographic Wikipedia articles where available. The appendix of the track description \citep{Ekstrand2022-ma} provides full details of the gender attribute, but they started with 20+ gender identities from Wikidata, collapsed transgender identities (treating trans men as men and trans women as women), and folding remaining gender identities into a third category; this resulted in ``male'', ``female'', ``third'' (``nonbinary'' in 2022), and ``unknown.'' This has the benefits of reducing combinatorial explosion and the number of groups with very few representatives, making the encoding more computationally practical. One downside of this approach is that it may obscure discrimination against binary transgender people specifically.

Our second recommendation is to document precisely how gender labels were obtained, whatever schema they use; prior work demonstrates that many datasets do not justify where they obtain the data nor the schema of data labels \citep{Scheuerman2020-cd}. This recommendation applies to both data obtained from existing sources, including public datasets, and new datasets created for particular projects. Such documentation should be reported in relevant publications and can also be a part of dataset documentation such as a datasheet \citep{Gebru2018-ml}. This document should document the schema used, the source of the data (such as self-identification or expert annotation), the construct of gender recorded (e.g. gender identity versus gender expression), and the principles used to determine gender when it is not self-identification. In specifying the schema, the documentation should also describe the options given to respondents, and if various instruments or interfaces limited options to the male/female gender binary. When working with existing data sets, such information may not be immediately obvious, but researchers and practitioners alike should perform due diligence to understand how gender was collected and recorded before working with the data. Documenting this information can serve as a community benefit for other researchers who seek to build on their results and/or work with the same data. When data is obtained from an intermediary, both the intermediary and the intermediary's source of gender data should be identified. In many of the papers we coded, the paper was not explicit about the source of gender data, and we had to infer the source from context, background assumptions, or other resources.

\ah{Our third recommendation is to consider greater gender diversity in one's data sample, especially when conducting small-n qualitative studies or user studies in which gender may be a significant factor for understanding results. We found that only 10 of our 73 papers which used a gender variable acknowledged non-binary gender, or provided a third option. None, however, positively affirmed a non-binary gender option. Therefore, it would be highly advisable that non-binary people are explicitly recruited for studies in which gender could be a key variable for both the auditing of a system, or for user studies which evaluate a system.}

Finally, when constructing new data sets for either research or application purposes, we recommend collection and curation that is thoughtful and respectful towards different gender identities\me{, as well as taking into account that} \ah{there is a danger in collecting demographic information in and of itself, as such information may make reductionist assumptions about identity \citep{hanna2020towards}, or be used in a way that violates privacy \citep{Keyes2022-ri}.} Self-identification is the best way to obtain gender data, as it most fully respects individual autonomy and self-determination, and it should be obtained through inclusive means. The HCI Gender Guidelines \citep{Scheuerman2020-dq} provides guidance for how to design gender-inclusive survey fields to obtain gender information from respondents. Expert annotation can be legitimate, but should be done in a way that respects peoples’ right to self-identify, along with their right to be excluded entirely. The Program for Cooperative Cataloging established a task force to produce recommendations for how to record the gender identities of book authors in library name authority records \citep{Billey2016-bb}, whose report provides explicit guidance about the type of inferences that should or should not be used when recording author information (when an author does not state their gender identity, the recommendations allow inference from clear indications in sources close to the author, such as the choice of gendered pronouns in an author’s own biography, but not from names or photographs). The relevant data field is also explicitly defined as recording an author’s gender identity \citep{Library_of_Congress_Network_Development2022-uf}.

\subsection{Research Needed}
Our systematic review and the recommendations we draw from it and relevant literature and guidance in adjacent fields are by no means the last word on the use and misuse of gender in information access. Further research is needed to identify and assess the various impacts of use-of-gender decisions. There are also open practical challenges: for example, while there is important work on measuring fairness beyond binaries \citep{Raj2022-js, Zehlike2022-cz}, it is not easy to deal computationally with rich notions of gender that may be multidimensional, combinatorially large, and have categories with relatively few members. When it is appropriate to use gender — for example, in audits for discrimination — the details of how to ethically, respectfully, and practically collect, store, document, analyze, and present rich notions of gender remain to be worked out.

There is also space to carry out similar analyses to understand how gender is being used in other fields such as natural language processing or data mining, and to document the use of gender in deployed industrial systems that are not yet described in the public research literature.

\section{Conclusion}
\label{sec:conclusion}
Gender is a complex and multifaceted construct that is often connected with important aspects of a person’s identity. A review of published literature reveals a variety of goals for which gender is employed. Pursuing gender equity in the effects of information access systems is an important goal, but this needs to be done thoughtfully and in a manner that respects the rights and identities of the people involved. Sometimes, gender should not be used; in other cases, it should be used but with due care and attention to the complexity of gender. This also needs to be accompanied with clear discussions of what, precisely, has been done, why, and limitations that arise from the chosen approach.

Our aim with this paper has been to provide an understanding of the current state of research practice and pointers to further reading to understand gender as it is currently understood, to serve as a foundation for robust, rigorous, and respectful investigations of how information access systems can avoid reproducing gender-related harms and can effectively serve users, content creators, and information subjects of all genders.

\begin{acks}
This work was supported by the National Science Foundation under Grant No. 17-51278.
\end{acks}

\bibliographystyle{ACM-Reference-Format}
\bibliography{chiir-gender}

\end{document}